\documentclass[twocolumn]{aastex701}

\usepackage[utf8]{inputenc}

\usepackage{graphicx}
\usepackage{txfonts}
\usepackage{lipsum}
\usepackage[T1]{fontenc}

\shorttitle{A Radio Search for SPI in TOI-540 and SPECULOOS-3}
\shortauthors{Ortiz Ceballos et al.}

\graphicspath{{./}{figures/}}

\begin{document}

\title{A Radio Search for Star-Planet Interaction in TOI-540 and SPECULOOS-3}

\author[0000-0003-3455-8814,gname=Kevin,sname=Ortiz Ceballos]{Kevin N. Ortiz Ceballos}
\affiliation{Center for Astrophysics ${\rm \mid}$ Harvard {\rm \&} Smithsonian, 60 Garden St, Cambridge, MA 02138, USA}
\email{kortizceballos@cfa.harvard.edu}

\author[0000-0001-7007-6295]{Yvette Cendes}
\affiliation{Department of Physics, University of Oregon, Eugene, OR 97403, USA}
\email{yncendes@uoregon.edu}

\author[0000-0002-9392-9681]{Edo Berger}
\affiliation{Center for Astrophysics ${\rm \mid}$ Harvard {\rm \&} Smithsonian, 60 Garden St, Cambridge, MA 02138, USA}
\email{eberger@cfa.harvard.edu}

\begin{abstract}
We present the first targeted centimeter-band radio observations of two recently-discovered exoplanet systems that are prime candidates for magnetic star-planet interaction (SPI): TOI-540 and SPECULOOS-3. The targets were selected due to the small orbital separation of their known planets, as well as for indications of stellar magnetic activity, given that for SPI radio emission may be strongest when a sufficiently magnetized star hosts a close-in planet. The deep, multi-epoch Very Large Array (SPECULOOS-3) and MeerKAT (TOI-540) observations yield non-detections, with $3\sigma$ limits of $\lesssim 7.5$ $\mu$Jy ($4-8$ GHz) and $\lesssim 30-80$ $\mu$Jy ($0.8-1.7$ GHz), respectively. For SPECULOOS-3 b we rule out observable SPI for most of its orbit, while for TOI-540 b we sample a narrower range, around planetary transit. We model possible planetary magnetic field strength constraints for both systems, and conclude that our observations are sensitive enough to sample SPI emission in these systems if present and directed at us, even for a planetary field of only $\sim 1$ G.

\end{abstract}


\section{Introduction} 
\label{sec:intro}

The magnetic interaction of an exoplanet with its host star can result in observable indicators that trace the planetary magnetic field properties. Planet-induced modulations in stellar chromospheric lines \citep{shkolnik_evidence_2003} and X-ray emission \citep{acharya_x-ray_2023}, as well as measurements of evaporation and bow shocks in transiting exoplanet atmospheres \citep{schreyer_using_2023} have been used to estimate planetary field strength and behavior. Induced stellar flaring correlated with planetary orbital periods has also been claimed as a potential physical mechanism of interaction between a host star and its planets \citep{ilin_planetary_2024, whitsett_induced-flare_2025, ilin_close-planet_2025}. Collectively, these processes are referred to as star-planet interactions (SPI).

Measuring planetary properties from observations of such processes is model-dependent and indirect. However, some SPI processes can induce radio emission, as seen in the Solar wind-driven, radio-bright aurorae of the solar system planets \citep{zarka_auroral_1992}. Radio observations offer an opportunity to determine the magnetic field strength (and its time-dependent behavior) of an exoplanet, because SPI leads to radio emission via the Electron Cyclotron Maser Instability (ECMI; e.g., \citealt{callingham_radio_2024}). ECMI emission occurs when energetic electrons are accelerated in the magnetic field lines of a quiescent magnetosphere, and is observed as highly circularly polarized emission \citep{treumann_electroncyclotron_2006}. ECMI is expected to exhibit a cutoff frequency that is directly proportional to the magnetic field strength ($B$), $\nu = 2.8 B$ MHz.

Magnetic SPI relies physically on the differential motion of a planet with respect to the large-scale magnetic field of its host star, in the presence of stellar wind \citep{strugarek_introduction_2025}. This interaction can cause observable radio emission in two ways: sub-Alfvénic interaction or stellar wind-magnetosphere interaction \citep{callingham_radio_2024}. In the sub-Alfvénic scenario, planets in close-in orbits are immersed in flowing magnetized plasma from the stellar wind, becoming obstacles that interact with the plasma flow and generate waves. When the orbit is close enough to lie within the Alfvén surface, energy is transported back to the star along these waves and can be observed as radio emission directly proportional to the planetary magnetic field strength \citep{saur_magnetic_2013}, similar to the ``Jupiter-Io'' effect in which bright periodic radio emission and auroral activity is observed from Jupiter in phase with the orbit of Io \citep{zarka_plasma_2007}. In stellar-wind magnetosphere interaction, radio emission is produced by the planet itself, driven by magnetic reconnection of the planet's magnetic field lines on the planetary night-side \citep{lanza_star-planet_2013, strugarek_moves_2022}. This mechanism is also known as the ``force-free'' or ``stretch-and-break'' model \citep{callingham_radio_2024}; we use the latter term here. In both scenarios, ECMI emission coming from SPI is highly directional \citep{kavanagh_hunting_2023}.

A principal challenge for radio observations of SPI is the limited number of systems within the optimal parameter space and at distances suitable for detection with current instruments, underscoring the importance of studying the nearest targets.  Moreover, radial velocity and transit searches exhibit a selection bias against magnetically active stars \citep{poppenhaeger_correlation_2011}. At GHz frequencies, polarized radio emission from stellar systems may arise from either ECMI or gyrosynchrotron processes, such that unambiguous identification of SPI requires detection of periodic emission in phase with the planetary orbital period. Radio emission from Proxima Centauri has been tentatively attributed to SPI based on such an orbital phase dependence \citep{perez-torres_monitoring_2021}. GHz emission has also been detected from the nearby Epsilon Eridani system \citep{bastian_radio_2018}, although the 7.6 year orbital period of its known planet precludes a comparable periodicity analysis. Periodic bursting emission from the M dwarf system YZ Ceti at $2-4$ GHz has been shown to coincide with the orbital phase of its innermost planet ($P_{\mathrm{orb}}\sim2$ d) and used to estimate its magnetic field \citep{pineda_coherent_2023,trigilio_star-planet_2023,pineda_evaluating_2025}.
We recently conducted a targeted survey of 77 planetary systems within 17.5 pc using the Karl G. Jansky Very Large Array (VLA); it resulted in the detection of magnetic activity from GJ 3323, a nearby M dwarf system with two close-in planets, one within the stellar Alfvén surface \citep{ortiz_ceballos_volume-limited_2024}. 

At lower frequencies, wide-field surveys such as the LOFAR Two-metre Sky Survey \citep[LoTSS; ][]{shimwell_lofar_2017,shimwell_lofar_2019} and its circularly polarized release V-LoTSS \citep{callingham_v-lotss_2023} have found coherent 150 MHz radio emission attributed to ECMI across the M dwarf main sequence \citep{callingham_population_2021}. 
Such emission, however, could be produced by intrinsic stellar processes and its detection is not in itself indicative of SPI.  For example, a detection of the M dwarf GJ 1151 in LoTSS was attributed to SPI emission following energy-scaling arguments \citep{vedantham_coherent_2020}, but follow-up radial velocity searches found only a small, long-period planetary companion incompatible with the proposed SPI scenario \citep{blanco-pozo_carmenes_2023}.

Here, we report the results of the first dedicated radio observations of two recently-discovered exoplanet systems selected for their potential of exhibiting detectable radio SPI, using the VLA and MeerKAT. These systems are compelling targets for SPI since they are nearby, host close-in planets, and have significant differential velocity between the planetary orbit and the rotation of the stellar magnetosphere.  We introduce the targets in \S\ref{sec:targets}, present the observations and and results in \S\ref{sec:observations}, and interpret the results in the context of SPI in \S\ref{sec:discussion}.

\begin{figure}[t!]
    \centering
    \vspace{-3mm}
    \includegraphics[width=\columnwidth]{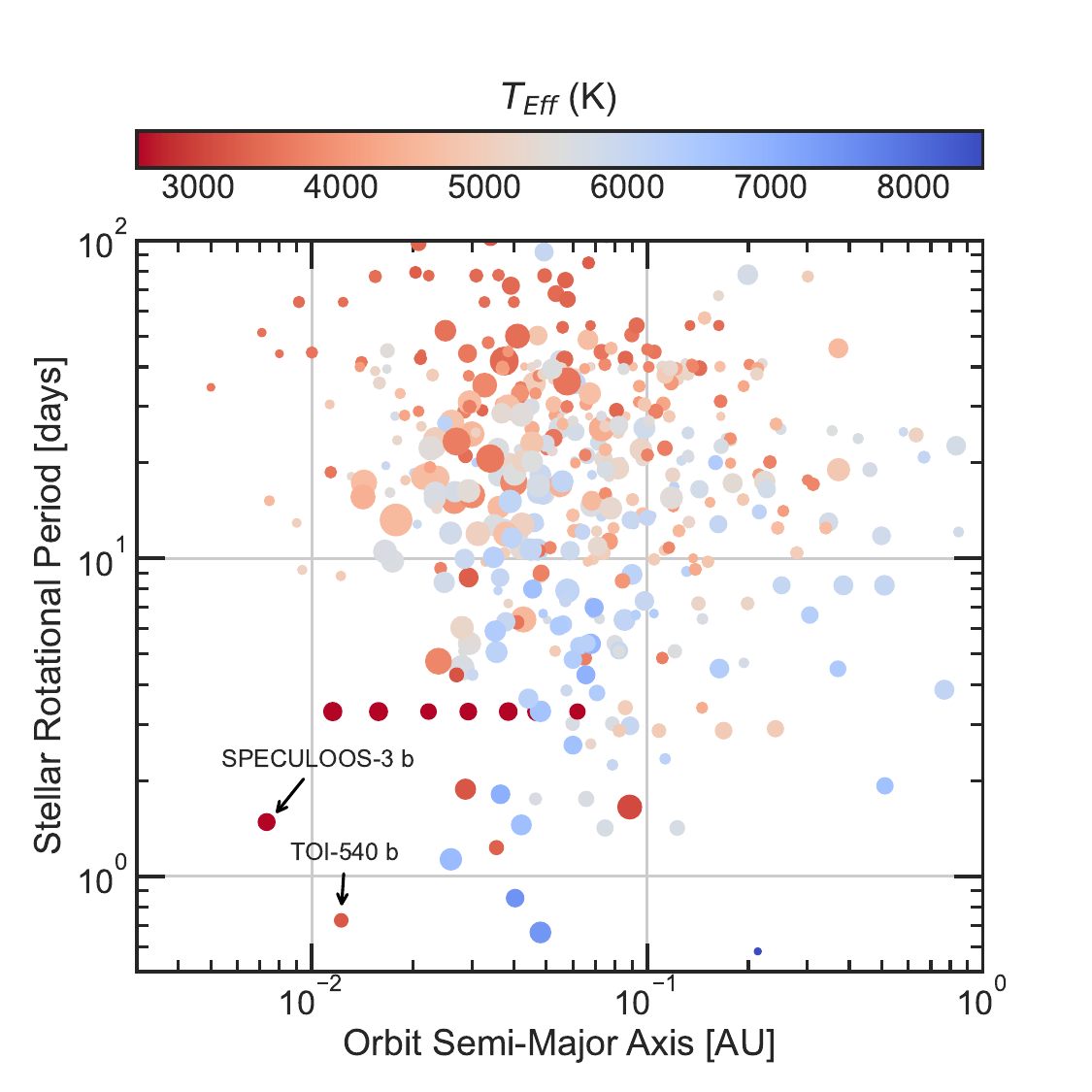}
    \vspace{-5mm}
    \caption{Stellar rotation period ($P_{\rm rot}$) versus orbital semi-major axis ($a$) for all known transiting exoplanets with $a\le 1$ AU and $P_{\rm rot}\le 100$ d, from the NASA Exoplanet Archive. Color denotes stellar effective temperature ($T_{\mathrm{eff}}$) and marker size is proportional to the planet to stellar radius ratio ($R_p/R_s$).  The two targets in this study, TOI-540 and SPECULOOS-3, are highlighted as the two systems in the archive with the smallest semi-major axes and shortest stellar rotation periods when choosing systems with known planetary radii. We note that the TRAPPIST-1 system is shown with the updated $P_{\rm rot} = 3.3$ d from \citet{luger_seven-planet_2017, vida_frequent_2017}.}
    \label{fig:rot}
\end{figure}

\section{Targets}\label{sec:targets}

We selected the two targets for this study because they were discovered recently \citep{ment_toi_2021, gillon_detection_2024}, and stand out from the sample of exoplanet systems in two key indicators for potential SPI activity: (i) both have a short stellar rotation period, as an indicator for stronger magnetic activity; and (ii) they have a small orbital separation, which can enable stronger SPI. We show these properties in relation to the sample of transiting exoplanets\footnote{From the NASA Exoplanet Archive, as of 2025 October 21; \href{https://exoplanetarchive.ipac.caltech.edu/}{https://exoplanetarchive.ipac.caltech.edu/}} in Figure~\ref{fig:rot}, which highlights the unique combination of rapid stellar rotation ($\lesssim 1.5$ d) and small semi-major axis ($\lesssim 0.012$ AU) for these two systems. Below, we describe each target in detail.

\smallskip
\noindent \textit{SPECULOOS-3.} This system consists of a rocky transiting planet ($R_p = 0.977 \pm 0.022\ R_{\oplus}$, $a = 0.007330 \pm 0.000055$ AU, $P_\mathrm{orb} = 0.71912603 \pm 0.00000057$ d), hosted by a M6.5 star ($d = 16.750 \pm 0.012$ pc. $P_\mathrm{rot} = 1.48 \pm 0.14$ d; \citealt{gillon_detection_2024}). The remarkably short planet-star separation ($\sim13$ stellar radii) results in an orbital period of only 17 hours; this not only increases SPI power, but also makes more of the orbital phase accessible with continuous radio observations. Furthermore, the host star's late spectral type places it in the ultracool dwarf (UCD) regime; such stars can be radio-bright on their own (e.g., \citealt{williams_radio_2018}) regardless of known companions. An age constraint of $6.6_{-2.4}^{+1.8}$ Gyr is determined from kinematic comparison with nearby co-moving stars \citep{gillon_detection_2024}. 

\smallskip
\noindent \textit{TOI-540.} This system consists of a rocky transiting exoplanet ($R_p = 0.903\pm 0.052\ R_{\oplus}$, $a = 0.01223 \pm 0.00036$ AU, $P_\mathrm{orb} = 1.2391491 \pm 0.0000017$ d) hosted by an M5 star ($d = 14.0$ pc. $P_\mathrm{rot} = 0.72610 \pm 0.00039$ d; \citealt{ment_toi_2021}).
Based on gyrochronological relations for stars later than spectral type M4, the star has an estimated age of $\sim 0.7$ Gyr \citep{engle_living_2023}. The planet's close orbital separation corresponds to 14 stellar radii from the host star, within the sub-Alfvénic regime.
Two different studies find statistically significant clustering of flares in-phase with the orbital period of the planet, potentially suggestive of SPI activity \citep{ilin_planetary_2024, whitsett_induced-flare_2025}.

\section{Observations and Results} 
\label{sec:observations}

\begin{table}
    \hspace*{-0.65cm}
    \begin{tabular}{ccccc}
    \toprule
    Date & Band & Synth.~Beam & Duration & RMS \\
         &      &             & (hr)     & ($\mu$Jy) \\
    \hline
    \multicolumn{5}{c}{SPECULOOS-3 -- VLA} \\ 
    \hline
    2024-07-20 & C & $0.98'' \times 0.89''$ & 1 & 4.5 \\
    2024-08-07 & L & $3.85'' \times 3.62''$ & 1 & 20.7 \\
    2025-07-30 & C & $3.57'' \times 3.05''$ & 4 & 2.5 \\
    2025-08-07 & C & $3.39'' \times 3.03''$ & 4 & 2.4 \\
    2025-08-17 & C & $3.26'' \times 3.01''$ & 4 & 2.3 \\
    2025-08-18 & C & $3.20'' \times 2.92''$ & 4 & 2.6 \\
    \hline
    \multicolumn{5}{c}{TOI-540 -- MeerKAT} \\ 
    \hline
    2025-05-25 & L & $11.9'' \times 11.8''$ & 1 & 27.6 \\
    2025-05-30 & L & $13.0'' \times 9.58''$ & 1 & 21.1 \\
    2025-06-01 & L & $8.13'' \times 7.63''$ & 1 & 12.4 \\
    2025-06-15 & L & $12.1'' \times 11.6''$ & 1 & 22.8 \\
    2025-06-16 & L & $7.48'' \times 7.36''$ & 1 & 10.6 \\
    \hline
    \end{tabular}
    \caption{Summary of radio observations presented in this work.}
    \label{tab:obslog}
\end{table}

\subsection{SPECULOOS-3}

We carried out two 1-hr exploratory observations of SPECULOOS-3 with the VLA in L-band ($1-2$ GHz) and C-band ($4-8$ GHz) through Director's Discretionary Time in July and August 2024 (Project ID: 24A-477, PI: Ortiz Ceballos; Table~\ref{tab:obslog}), shortly after the discovery of the system was announced. We subsequently obtained four 4-hr observations with the VLA in C-band in July to August 2025 (Project ID: VLA/25A-296, PI: Ortiz Ceballos; Table~\ref{tab:obslog}) covering different portions of the planetary orbital phase. All VLA observations were carried out in the standard continuum mode. In all cases, the bandpass calibrator was 3C48 and the phase calibrator was J2052+3635. 

\begin{figure*}[h]
    \centering
    \includegraphics[width=0.95\textwidth]{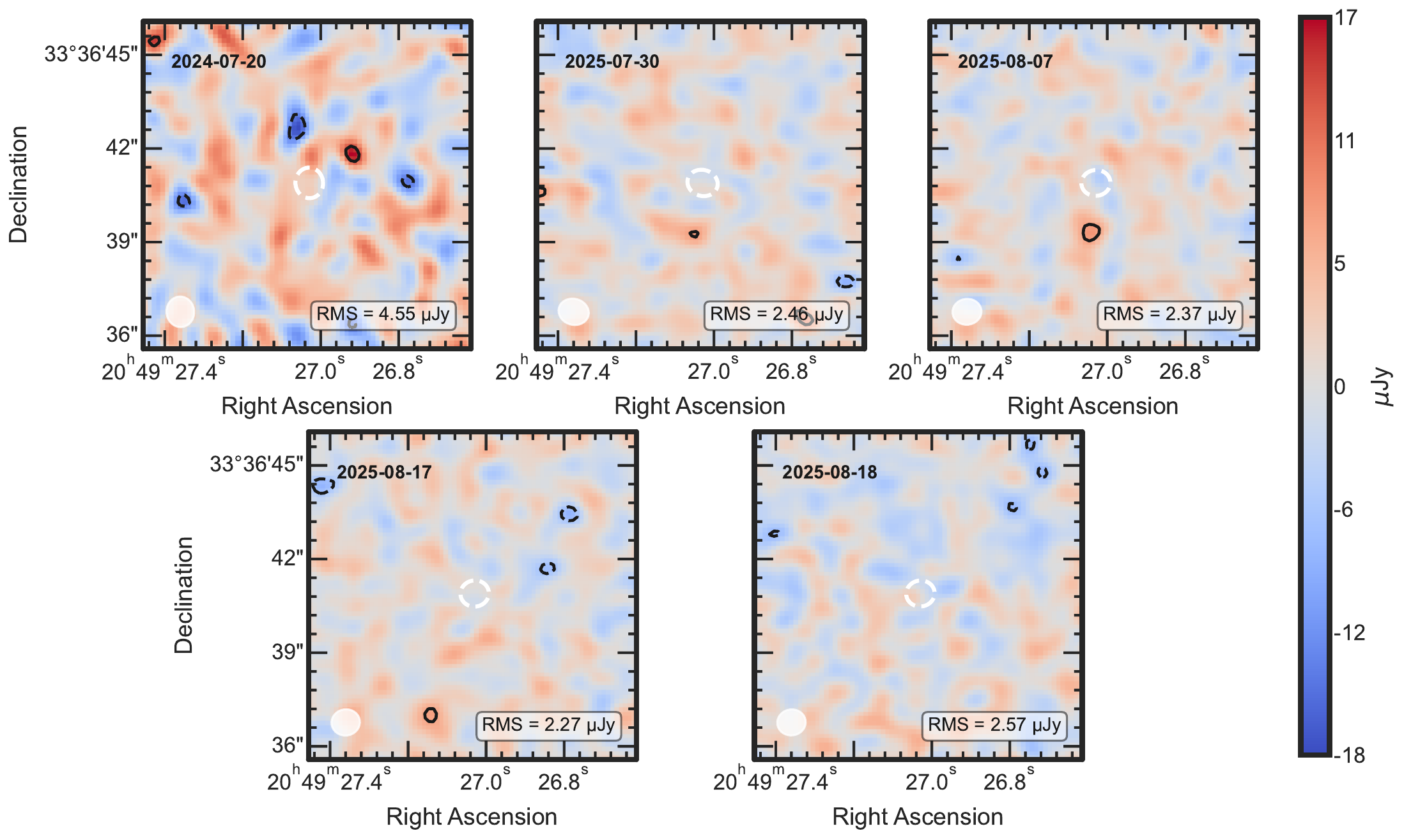}
    \caption{VLA C-band observations of SPECULOOS-3. The upper left panel is a 1-hr observation, while the remaining four observations are 4 hours each.  The synthesized beam is shown in the bottom left corner of each panel, and the outlined beam in the center of each panel shows the \textit{Gaia} proper-motion corrected coordinates of SPECULOOS-3 during each observation. Contours trace -3$\sigma$, 3$\sigma$, 5$\sigma$ and 7$\sigma$.}
    \label{fig:speculoos_images}
\end{figure*}

The data were processed with the standard NRAO VLA pipeline included in the Common Astronomy Software Applications \citep{casa_team_casa_2022}. The 2024 observations were processed with CASA version 6.5.4, and the 2025 observations with CASA version 6.6.1. All C-band observations were obtained as image products from the NRAO Science Reference Data Products (SRDP) pipeline. The L-band observation was not processed by the SRDP program, so we used the pipeline-calibrated visibilities and imaged it directly with CASA \verb|tclean|. The C-band images are shown in Figure~\ref{fig:speculoos_images}, and the L-band image in Figure~\ref{fig:speculoos_Lband}. SPECULOOS-3 was not detected in any of the observations, with resulting $3\sigma$ flux density upper limits of $\approx 7.5$ $\mu$Jy in C-band and $\approx 20$ $\mu$Jy in L-band.

We also note that the location of SPECULOOS-3 was covered in the VLA Sky Survey \citep[VLASS; ][]{lacy_karl_2020} in S-band ($2-4$ GHz), with resulting shallower $3\sigma$ upper limits\footnote{We accessed VLASS using the Quicklook cutouts available via the Canadian Astronomy Data Centre (\href{https://www.cadc-ccda.hia-iha.nrc-cnrc.gc.ca/en/search/?collection=VLASS}{https://www.cadc-ccda.hia-iha.nrc- cnrc.gc.ca/en/search/?collection=VLASS}).} of $\approx 380$ $\mu$Jy on 2019 June 5, $\approx 340$ $\mu$Jy on 2021 October 13, and $\approx 340$ $\mu$Jy on 2024 July 1.

\begin{figure*}[h]
    \centering
    \includegraphics[width=0.95\textwidth]{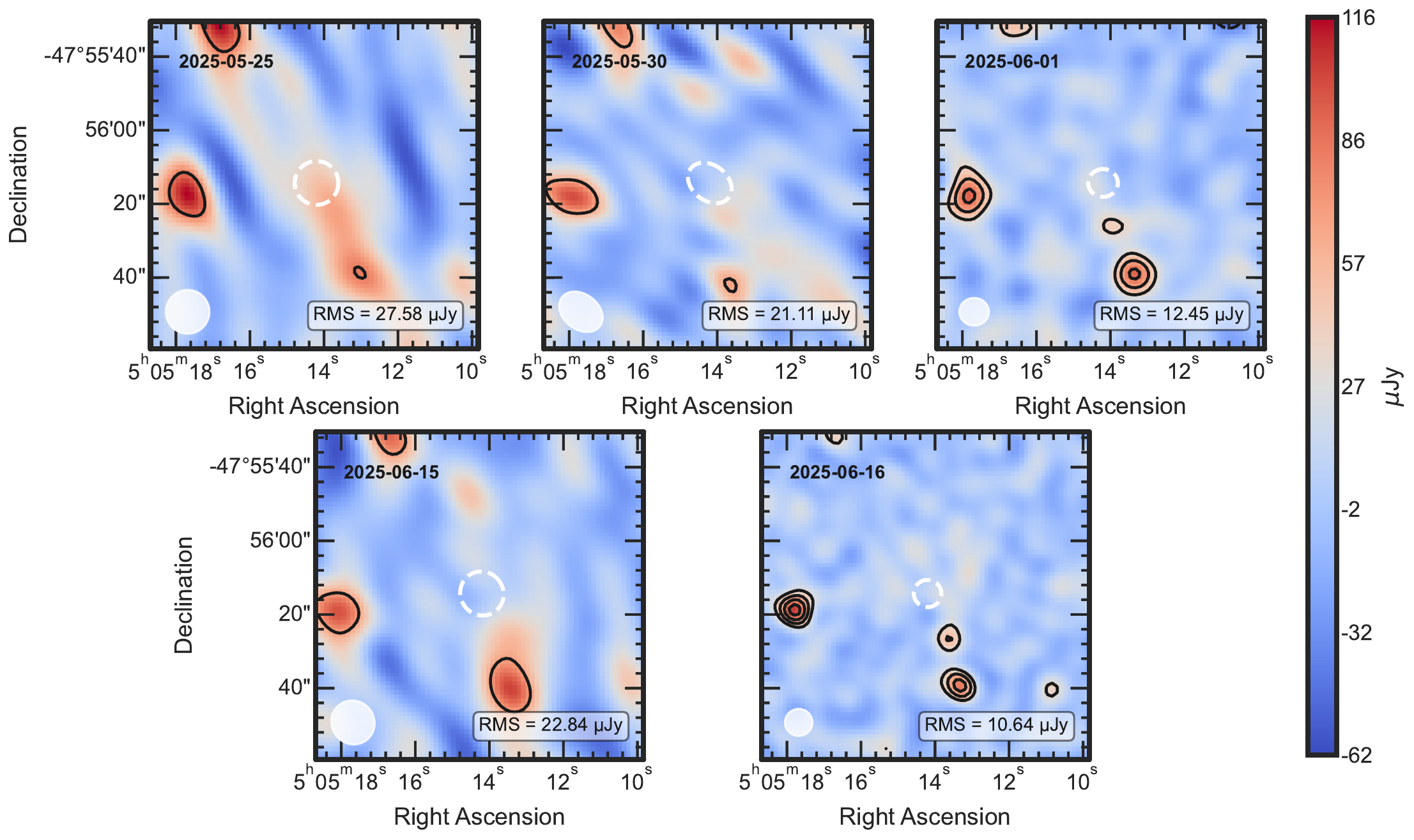}
    \caption{Same as Figure~\ref{fig:speculoos_images} but for MeerKAT L-band observations of TOI-540.}
    \label{fig:toi540_images}
\end{figure*}

\subsection{TOI-540}

We observed TOI-540 in five epochs spanning May to June 2025 with MeerKAT (Program ID: SCI-20241102-KO-01, PI: Ortiz Ceballos) in the standard L-band observing mode, with 4096 spectral channels across a $\approx 1$ GHz bandwidth window ($0.8-1.7$ GHz; Table~\ref{tab:obslog}). For all epochs we used J0408$-$6545 as the bandpass calibrator, and J0440$-$4333 as the phase calibrator. The observations were processed by the South African Radio Astronomy Observatory Science Data Processor\footnote{\href{https://skaafrica.atlassian.net/wiki/spaces/ESDKB/pages/338723406/}{https://skaafrica.atlassian.net/wiki/spaces/ESDKB/pages/338723406/}} (SDP). We use the calibrated, primary-beam corrected mean continuum images provided by the SDP. TOI-540 was not detected in any of the observations, with resulting $3\sigma$ upper limits of $\approx 30-80$ $\mu$Jy; see Figure \ref{fig:toi540_images} and Table~\ref{tab:obslog}.

\begin{figure}
   \centering
   \includegraphics[width=\columnwidth]{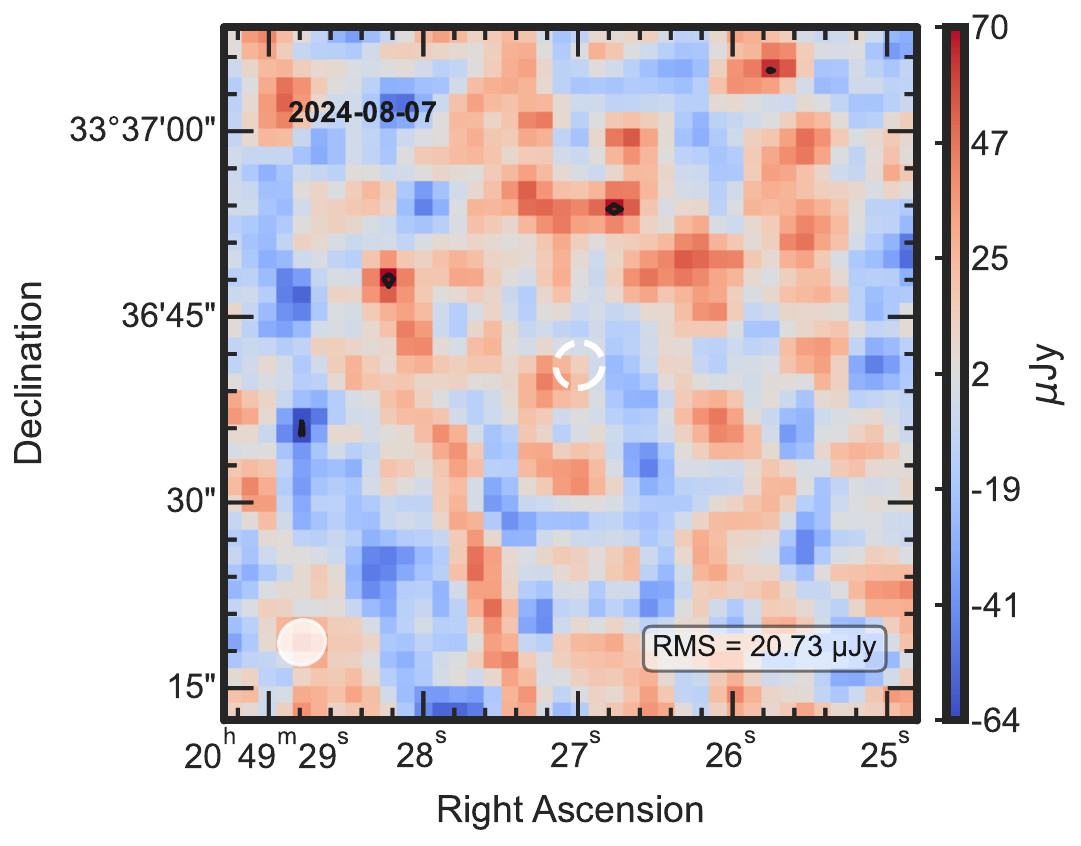}
   \caption{Same as Figure~\ref{fig:speculoos_images} but for the L-band VLA observation of SPECULOOS-3.}
   \label{fig:speculoos_Lband}
\end{figure}

\section{Discussion} 
\label{sec:discussion}

\begin{figure*}
    \centering
    \includegraphics[width=\textwidth]{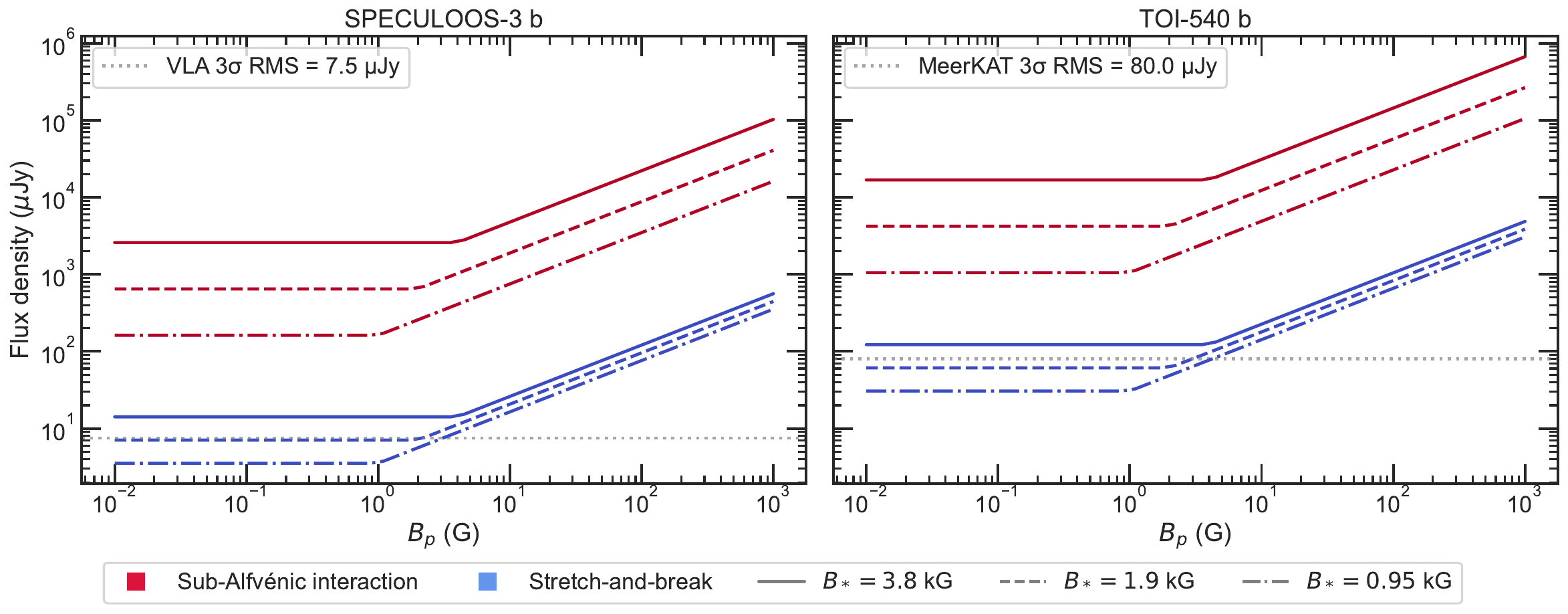}
    \caption{Estimated radio flux density from SPI for the sub-Alfvénic interaction (blue) and stretch-and-break (red) scenarios. We show three different stellar magnetic field strength assumptions. The flat part of each curve represents the regime where the planets are effectively unmagnetized. For both targets our observations reach the sensitivity limits needed to infer a sub-Gauss planetary field if SPI is present during the observations. The dotted horizontal lines show our $3\sigma$ RMS non-detection sensitivity.}
    \label{fig:energy}
\end{figure*}

We consider two physical scenarios in which magnetic SPI can induce observable radio signals from SPECULOOS-3 and TOI-540; namely, sub-Alfvénic interaction, and stretch-and-break reconnection. In both scenarios, electrons are accelerated in the star-planet system, producing ECMI emission \citep[see][for a review]{treumann_electroncyclotron_2006}. The resulting emission should vary in phase with the planetary orbit, making identification as radio SPI (as opposed to radio flares or activity from the star) possible.

The Jupiter-Io system serves as a nearby laboratory for sub-Alfvénic magnetic interaction \citep{zarka_auroral_1998}. When the companion orbits within the Alfvén surface of the host, the companion's motion perturbs the host's magnetic field, creating Alfvén waves, which are transverse plasma waves generated mechanically by the planet's motion \citep{neubauer_nonlinear_1980}. These waves propagate back to the host star via its magnetic field lines, transporting energy that accelerates electrons. When the accelerated electrons have sufficiently large pitch angles, they create an observable ECMI effect in the star's magnetic field \citep{callingham_radio_2024}. 

In the stretch-and-break model, the planetary magnetic field is stretched by the ambient stellar wind magnetic field as the planet moves along its orbit across the stellar wind, forcing planetary field lines that face the star into reconnecting on the planetary night-side \citep{lanza_star-planet_2013, strugarek_moves_2022}. The energy released by the reconnection creates an auroral cone of ECMI emission that can be observed as radio SPI.

To model the expected radio fluxes from the sub-Alfvénic interaction and stretch-and-break scenarios for SPECULOOS-3 and TOI-540 we follow an approach similar to \citet{pineda_coherent_2023}, using their equations 1 through 4. We limit ourselves to the case of a stellar magnetosphere with a radial isothermal wind, due to the uncertain nature of the magnetic field and wind environments for SPECULOOS-3 and TOI-540.
We assume a plasma environment similar to the Io-Jupiter system ($m_\mathrm{ion} = 22$ amu, $n \sim 2000 \mathrm{~cm}^{-3}$; as in e.g., \citealt{pineda_deep_2018}).
We take the surface average stellar magnetic field strength of 3.8 kG for TOI-540 from \citet{ilin_planetary_2024}, and use their equations 4 and 5 to calculate an equivalent field strength of 3.8 kG for SPECULOOS-3.
We then also consider scalings of the stellar field to 50\% and 25\% of this value (1.9 kG and 0.95 kG, respectively) to produce more conservative radio flux predictions, since these magnetic field values can be overestimated.
For example, \citet{ilin_planetary_2024} find a field strength of $1100 \pm 200$ G for YZ Ceti, while Zeeman Doppler Imaging maps find a surface average stellar magnetic field strength of just $\sim 225$ G for the star \citep{pineda_evaluating_2025}. The earlier-type fast rotators StKM 1-1262 (M0 type, Prot = 1.24 d) and V374 Peg (M4 type, Prot = 0.4455 d) have measured surface average fields of 300 and 800 G respectively, and exhibit radio emission \citep{bellotti_magnetic_2025}. While the kilo-Gauss field strength estimates are reasonable given observations of other fast-rotating late-type M-dwarfs \citep{shulyak_strong_2017}, this depends on the stellar magnetic field structure (strongly dipolar vs multipolar), unknown for our target stars.

Once a stellar magnetic field strength is assumed, the stellar wind magnetic field at the location of the planet can be determined. We take the stellar field to decay radially as $B\propto 1/r^3$ out to a distance of 5 $R_*$, then as $B\propto 1/r^2$ after that, similar to \cite{pineda_evaluating_2025}.
For both systems, this leads to a stellar wind field strength of $\sim$4 G at the location of the planet, which reduces to $\sim$2 G and $\sim$1 G when assuming the 50\%- and 25\%- scaled fields.
For the geometric factor $\theta$, we assume $\pi/2$, which is an appropriate assumption for planets with $a \sim 0.01$ AU \citep[see Figure 9 of ][ where $\Theta = \pi/2 + \theta=\pi$]{saur_magnetic_2013}.
We use Equation 3 of \citet{pineda_coherent_2023} to estimate the size of the planetary magnetosphere. Finally, we assume a standard 1\% conversion efficiency to radio flux and a 1.6 sr beam solid angle \citep{turnpenney_exoplanet-induced_2018}, and consider appropriate bandwidths of 1 GHz for MeerKAT (TOI-540) and 4 GHz for the VLA (SPECULOOS-3).

The resulting predicted flux densities as a function of the unknown planetary magnetic field are shown in Figure~\ref{fig:energy}. We find that the sensitivity achieved in our observations, when compared to the model predictions, could in principle probe sub-Gauss (i.e. Earth-like) planetary magnetic fields. For the stretch-and-break scenario, our observations would be sensitive to SPI from a scenario where the empirically known radius of both planets is the only obstacle in the plasma flow. This means that the stretch-and-break model does not in this case require that the planet be magnetized. 
For the case of sub-Alfvénic interaction, SPECULOOS-3 and TOI-540 are interesting systems because their planets are close enough to their host stars to surpass the $0.1-0.3$ AU valley where SPI power is reduced to 0 due to alignment between the stellar wind magnetic field and the ambient plasma flow \citep{saur_magnetic_2013}.
Our observations are sensitive enough to reach the unmagnetized planet scenario for sub-Alfvénic interaction in both systems (corresponding to the flat part of the flux density lines in Figure \ref{fig:energy}), as long as the 3.8 kG surface averaged stellar magnetic field strength is assumed. The 50\% scaling brings the unmagnetized scenario right to our sensitivity limit for both targets, and the 25\% scaling places it outside of our detection threshold. 

\begin{figure}
\gridline{
    \fig{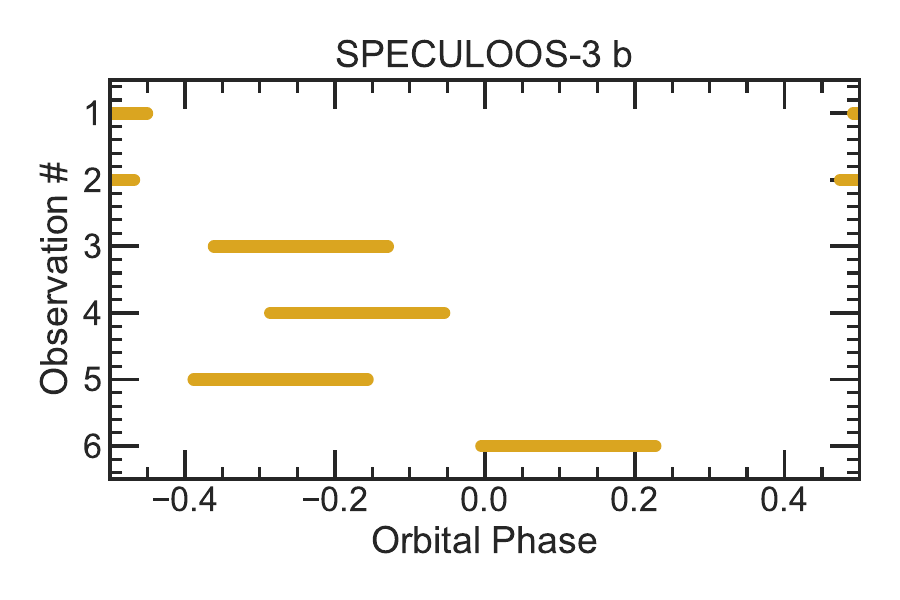}{0.95\columnwidth}{}
}
\vspace{-10mm}
\gridline{
    \fig{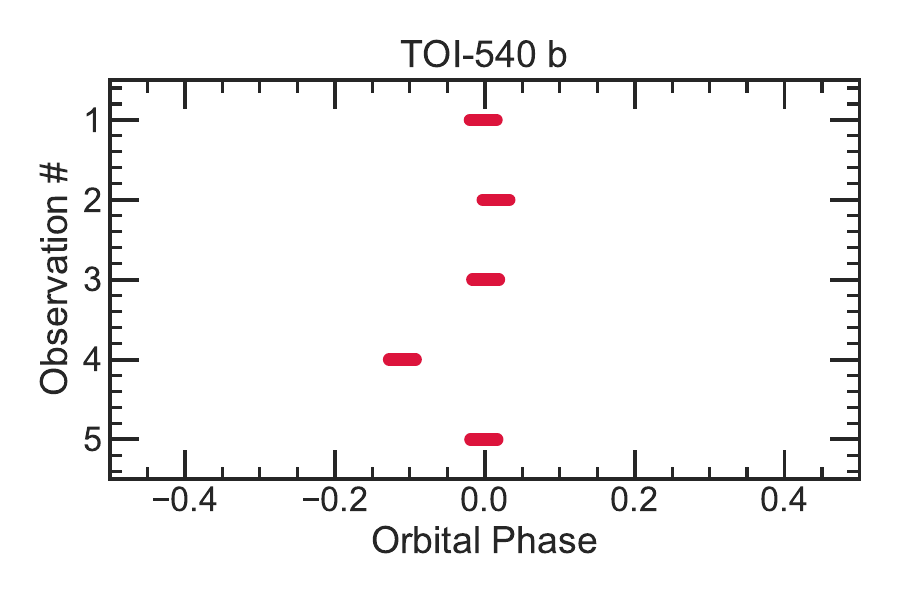}{0.95\columnwidth}{}
}
\vspace{-10mm}
\caption{Orbital phase coverage of our radio observations. The orbital parameters for SPECULOOS-3 b are from \citet{gillon_detection_2024}, and for TOI-540 b are from \citet{ment_toi_2021}.}
\label{fig:phases}
\end{figure}

\begin{figure*}
	\centering
	\includegraphics[width=\textwidth]{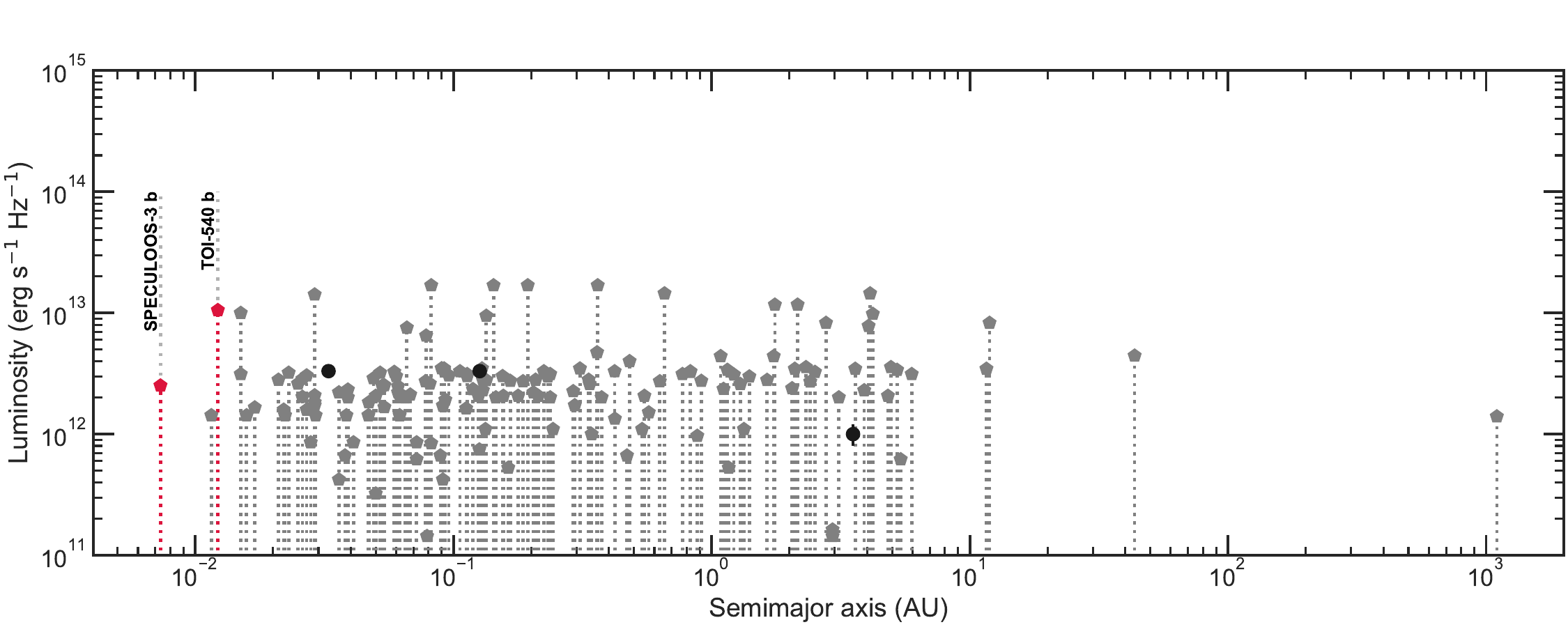}
	\caption{\label{fig:LumLimits} Luminosity upper limits as a function of orbital separation for SPECULOOS-3 b and TOI-540 b (red), compared to the sample (grey) from \cite{ortiz_ceballos_volume-limited_2024} and the literature measurements discussed therein \citep{bower_radio_2009, bastian_radio_2018, pineda_deep_2018, cendes_pilot_2021}. Upper limits are shown as markers with dotted lines, while the three circular markers in black correspond to, in order of increasing orbital separation, GJ 3323 b, GJ 3323 c, and Epsilon Eridani b.
    }
\end{figure*}

It is important to highlight the geometric limitations to the detectability of any SPI, which may be responsible for the non-detections of emission even if magnetic SPI was taking place in either system. In the Jupiter-Io system, emission is seen at quadrature \citep{bigg_influence_1964}. However, this orbital phase dependence is not exact, and will depend on variables such as the orbital inclination, projected spin-orbit angle and magnetic obliquity \citep{kavanagh_hunting_2023}. To calculate the phase coverage for both systems in the observations reported here, we used the reported transit epoch and orbital periods from the discovery papers \citep{gillon_detection_2024, ment_toi_2021} and propagated them to our observing epochs. We set phase = 0 as the time of transit center. We show the phase coverage of our observations in Figure \ref{fig:phases}. Our 4-hr observations of SPECULOOS-3, and the timing of the observations lead to a wide coverage of the orbital phase space, including the quadrature points. 
For TOI-540, the shorter 1-hr blocks, and the timing of the observations, resulted in mainly coverage of phase = 0.

We also note that sub-Alfvénic SPI emission is expected to be highly beamed, and thus the geometry of the system could make the emission undetectable to us. Without better observational constraints, it is difficult to predict how this geometric factor plays out in radio SPI. Both SPECULOOS-3 b and TOI-540 b are transiting exoplanets (and thus their orbital inclinations are known). M-dwarf targets of radio SPI searches have mostly been non-transiting planets otherwise \citep[e.g. ][]{turnpenney_exoplanet-induced_2018, perez-torres_monitoring_2021, pineda_coherent_2023}, with exceptions \citep[e.g. ][]{pena-monino_searching_2024}. The geometry of transiting systems can beam SPI towards Earth observers, but will only do so for a small portion of the orbital phase space \citep{kavanagh_hunting_2023}, underscoring the importance of sampling the orbital phase as completely as possible. This is a significant limitation of our results for the TOI-540 system, for which the observed epochs cluster around phase = 0.

Beyond emission from SPI, we note that some key indicators did point to both TOI-540 and SPECULOOS-3 as being magnetically active and thus potentially radio-bright objects in their own right: (i) both of their spectra show H$\alpha$ emission \citep{ment_toi_2021, gillon_detection_2024}, significantly correlated with detectable radio emission in UCDs \citep{pineda_panchromatic_2017}; (ii) their fast (0.7 and 1.48 d, respectively) rotation periods place them in the saturated regime of the rotation-activity relation \citep{wright_stellar_2018}; and (iii) TESS light curves show them to be flaring stars \citep{ment_toi_2021, gillon_detection_2024}. Additionally, TOI-540 is also detected in the soft X-rays at ${\rm log}(L_X) = 27.8$ erg s$^{-1}$ in the eRASS1 survey \citep{merloni_srgerosita_2024}. SPECULOOS-3 lies outside the eRASS1 field; it was also not matched with a source in the original RASS survey \citep{freund_stellar_2022}. However, the later spectral type of SPECULOOS-3 (M6.5) makes it only the second UCD known to host an exoplanet, after TRAPPIST-1 (spectral type M8). It has been speculated that radio-bright UCDs may host close companions that supply the electrons necessary for radio emission and that could even imprint time-variability \citep{kao_constraints_2019,williams_radio_2018}. While TRAPPIST-1 does not show detectable radio emission \citep{pineda_deep_2018}, the earlier type UCDs such as SPECULOOS-3 tend to display the highest radio luminosities of all UCDs \citep{williams_trends_2014,pineda_panchromatic_2017}. 

\section{Conclusions} 
\label{sec:conclusion}

We presented the first deep, targeted centimeter-band radio observations of two of the most promising radio SPI candidate systems to date, TOI-540 and SPECULOOS-3. The targets were chosen because of their expected stellar magnetic activity and close star-planet separation, which make potential SPI processes more likely to be detectable. We sampled most of the orbital phase of SPECULOOS-3, and obtained multiple epochs near planetary conjunction for TOI-540. We were able to place strict upper limits on the radio emission from each target: $F_\nu\lesssim 30$ $\mu$Jy for TOI-540 and $F_\nu\lesssim 7.5$ $\mu$Jy for SPECULOOS-3.

The wide orbital phase coverage in the observations of SPECULOOS-3 b, as well as the higher sensitivity and coverage in both L- and C-bands, allow a more confident conclusion that there is no radio SPI taking place in the system. For TOI-540, the limited phase coverage means that sharp SPI signals visible only at certain phases cannot be ruled out \citep{kavanagh_hunting_2023}.
We place our results in context with other nearby systems observed for SPI emission in Figure \ref{fig:LumLimits}.
Next-generation radio telescopes such as the Square Kilometer Array (SKA) and the Next Generation Very Large Array (ngVLA) will enable an order of magnitude more sensitive searches of these and similar systems, and may reveal the first confirmed radio SPI signals.

\begin{acknowledgments} 
The Berger Time-Domain Group at Harvard is supported by NSF and NASA grants. K.O. is supported by NSF Graduate Research Fellowship (GRFP), grant number DGE1745303, and the Ford Foundation Predoctoral Fellowship.
The National Radio Astronomy Observatory and Green Bank Observatory are facilities of the U.S. National Science Foundation operated under cooperative agreement by Associated Universities, Inc.
The MeerKAT telescope is operated by the South African Radio Astronomy Observatory, which is a facility of the National Research Foundation, an agency of the Department of Science and Innovation.
This research has made use of the NASA Exoplanet Archive, which is operated by the California Institute of Technology, under contract with the National Aeronautics and Space Administration under the Exoplanet Exploration Program.
This work has made use of data from the European Space Agency (ESA) mission {\it Gaia} (\url{https://www.cosmos.esa.int/gaia}), processed by the {\it Gaia} Data Processing and Analysis Consortium (DPAC, \url{https://www.cosmos.esa.int/web/gaia/dpac/consortium}). Funding for the DPAC has been provided by national institutions, in particular the institutions participating in the {\it Gaia} Multilateral Agreement.

\end{acknowledgments}

\bibliographystyle{aasjournalv7}

\end{document}